\documentclass[sn-mathphys,Numbered]{sn-jnl}

\usepackage{graphicx}%
\usepackage{multirow}%
\usepackage{amsmath,amssymb,amsfonts}%
\usepackage{amsthm}%
\usepackage{mathrsfs}%
\usepackage[title]{appendix}%
\usepackage{xcolor}%
\usepackage{textcomp}%
\usepackage{manyfoot}%
\usepackage{booktabs}%
\usepackage{algorithm}%
\usepackage{algorithmicx}%
\usepackage{algpseudocode}%
\usepackage{listings}%
\usepackage{makecell}

\usepackage{multirow}

\theoremstyle{thmstyleone}%
\theoremstyle{thmstyletwo}%

\theoremstyle{thmstylethree}%

\begin{document}

\title{MELEP: A Novel Predictive Measure of Transferability in Multi-Label ECG Diagnosis}


\author[1]{\fnm{Cuong V.} \sur{Nguyen}}\email{cuong.nv@vinuni.edu.vn}

\author[1]{\fnm{Hieu Minh} \sur{Duong}}\email{hieu.dm@vinuni.edu.vn}

\author*[1,2]{\fnm{Cuong D.} \sur{Do}}\email{cuong.dd@vinuni.edu.vn}

\affil[1]{\orgdiv{College of Engineering and Computer Science}, \orgname{VinUniversity}, \orgaddress{\city{Hanoi}, \country{Vietnam}}}

\affil[2]{\orgdiv{VinUni-Illinois Smart Health Center}, \orgname{VinUniversity}, \orgaddress{\city{Hanoi}, \country{Vietnam}}}


\abstract{
 In practical electrocardiography (ECG) interpretation, the scarcity of well-annotated data is a common challenge. Transfer learning techniques are valuable in such situations, yet the assessment of transferability has received limited attention. To tackle this issue, we introduce MELEP, which stands for \textit{Muti-label Expected Log of Empirical Predictions}, a measure designed to estimate the effectiveness of knowledge transfer from a pre-trained model to a downstream multi-label ECG diagnosis task.
 MELEP is generic, working with new target data with different label sets, and computationally efficient, requiring only a single forward pass through the pre-trained model.
 To the best of our knowledge, MELEP is the first transferability metric specifically designed for multi-label ECG classification problems.
 Our experiments show that MELEP can predict the performance of pre-trained convolutional and recurrent deep neural networks, on small and imbalanced ECG data.
 Specifically, we observed strong correlation coefficients (with absolute values exceeding 0.6 in most cases) between MELEP and the actual average F1 scores of the fine-tuned models.
 Our work highlights the potential of MELEP to expedite the selection of suitable pre-trained models for ECG diagnosis tasks, saving time and effort that would otherwise be spent on fine-tuning these models.
 }





\maketitle

\section{Introduction}
\label{sec:intro}

Electrocardiogram (ECG) signals are non-invasive and cost-effective tools for early detection and accurate diagnosis of heart-related disease, one of the leading causes of death worldwide. Early diagnosis and treatment can improve patient outcomes, making ECG signals essential for improving health and well-being. Recently, automatic ECG interpretation has gained significant popularity and witnessed remarkable progress.
This advancement can be attributed to the wide-scale digitization of ECG data and the evolution of deep learning techniques.
Notably, deep neural networks (DNN) have achieved classification performance on par with cardiologists, as demonstrated by Hannun et al. \cite{hannun2019cardiologist}, and Ribeiro et al. \cite{ribeiro2020automatic}.
These outstanding achievements have partly been due to the availability of extensive human-labeled datasets, consisting of 91,232 and 2,322,513 ECG recordings, respectively. 
However, ECG datasets used in practice are often much smaller, due to the expensive and time-consuming data collection and annotation process.
Consequently, it becomes challenging to achieve desirable results when training DNNs from scratch.
Transfer learning is often useful in such scenarios, resulting in improved performance \cite{headretraining2, kornblith2019better} and faster convergence \cite{he2019rethinking}.
Fortunately, there exists some large, publicly available ECG datasets, which enable DNNs to learn important latent features, then transfer the learned knowledge to our main task, typically with much less annotated data.
There are two most commonly used transfer learning techniques: head retraining \cite{headretraining1,headretraining2} and fine-tuning \cite{finetuning1, finetuning2}.
Both replace the top classification layer to match the number of target task's outputs; however, whereas the former freezes all feature extractor layers and only updates the top layer's parameters during training on the target dataset, the latter does not have such a constraint and makes all layers trainable.
Research suggested that fine-tuning leads to better performance \cite{leep, kornblith2019better,nguyen2024transfer}, thus it has been accepted as a de facto standard.

Given the effectiveness of fine-tuning, a new problem arises: how do we select the best pre-trained checkpoints among a large candidate pool?
A checkpoint is a model pre-trained on a source dataset, with a specific set of hyperparameter settings.
It is straightforward to actually do the fine-tuning and then select the top ones; however, this method is obviously expensive and difficult to scale.
Transferability estimation \cite{ammar2014automated, sinapov2015} aims to address the above bottleneck by developing a metric that indicates how effectively transfer learning can apply to the target task, ideally with minimal interaction with it.
Good estimation is likely to facilitate the checkpoint selection process.
In the domain of computer vision, several transferability measures were developed.
Tran et al. \cite{tran2019NCEscore} introduced negative conditional entropy between the source and target label sets.
Bao et al. \cite{bao2019hscore} proposed a transferability measure called H-score, which was based on solving a Maximal HGR Correlation problem \cite{hirschfeld1935connection,gebelein1941statistische,renyi1959measures}.
Nguyen et al. \cite{leep} and Huang et al. \cite{huang2022frustratingly} developed LEEP and TransRate, respectively, two efficient estimates with no expensive training on target tasks.
However, those measures only apply to multi-class classification problems, and thus cannot be directly applicable to multi-label tasks such as ECG diagnosis, in which a patient may suffer from more than one cardiovascular disease.

\vspace{2mm}
\textbf{Key Contributions}:

\begin{itemize}
 \item We propose MELEP, a transferability measure that can directly apply to multi-label classification problems in automatic ECG interpretation. 
 To the best of our knowledge, we are the first to develop such a measure to estimate the effectiveness of transfer learning for multi-label ECG.
 
 \item We conducted the first extensive experiment of transfer learning for 12-lead ECG data. We focused on small downstream datasets and covered a wide range of source checkpoints, which were produced from multiple source datasets and representatives of the two most popular DNN architectures for time-series analysis: convolutional and recurrent neural networks.
\end{itemize}

Our article is structured as follows: first, we provide the mathematical foundation behind MELEP and describe the intuition and its properties. 
Then four 12-lead ECG datasets and two DNN architectures are introduced, which build the backbone of our experiments.
We evaluate the ability of MELEP to predict the fine-tuning performance of a convolutional neural network by conducting extensive experiments with multiple checkpoints produced from pre-training the model on different source datasets.
To show the versatility of MELEP, we replicate the experiment with a recurrent neural network, affirming that its capability is not tied to a specific model architecture.
Next, we demonstrate the effectiveness of MELEP in a real-world scenario, which is selecting the best checkpoints among a group of pre-trained candidates.
Finally, we discuss some notable properties, extensions, and applications of MELEP and suggest promising directions for future study.

\section{Materials \& Methods}
\label{sec:methods}

\subsection{Multi-Label Expected Log of Empirical Predictions (MELEP)}
\label{subsec:melep}

Consider transfer learning from one multi-label classification task to another. 

Let:
\begin{itemize}
 \item $\Theta$ be the pre-trained model on the source task.
 \item $\mathcal{L}_s = \{0, 1, ..., \mathcal{Z} \! - \! 1\}$ be the source label set of size $|\mathcal{L}_s| = \mathcal{Z} $.
 \item $\mathcal{L}_t = \{0, 1, ..., \mathcal{Y} \! -1 \! \}$ be the target label set of size $|\mathcal{L}_t| = \mathcal{Y} $.
 \item $\mathcal{D} = \{ (x_1, \mathbf{y_1}), ..., (x_n, \mathbf{y_n}) \}$ be the target dataset of size $n$. $\mathbf{y_i}$ is a label vector of size $\mathcal{Y}$.
 \item $(y, z) \in \mathcal{L}_t \times \mathcal{L}_s$ be a pair of target-source labels taken from the two sets.
 \item $(t, s)$ be the values of $(y, z)$. In the ECG classification context, the label values are binary, so $(t,s) \in \{0, 1\} \times \{0, 1\}$.
\end{itemize}

then MELEP is computed as follows:

\begin{enumerate}
 \item Step 1: Compute the dummy label distributions of the target data over the source label set, denoted by a vector $\mathbf{\hat{y}}_i= \Theta(x_i)$ of size $\mathcal{Z}$, by forward passing each data point to the pre-trained model.

 \item Step 2: Consider each pair of target-source labels $(y, z)$. Let $\theta_{iz}$ denote the value of $\mathbf{\hat{y}}_i$ at the $z^{\text{th}}$ column, i.e. the predicted probability that the sample $x_i$ belongs to label $z$.
 
 \begin{enumerate}
 \item Compute its $2 \times 2$ empirical joint distribution matrix $\mathbf{\hat{P}_{yz}}(t,s)$, with value at row $t$ column $s$ is: 
 \begin{equation}
 \hat{P}_{yz}(t, s) = \frac{1}{n} \sum_{i: y_{iz}=t} (\theta_{iz})_s
 \end{equation}
 where $\sum_{i: y_{iz}=t}$ means we select all samples $x_i$ with the $z^{\text{th}}$ ground-truth label $y_{iz}$ equal to $t$. With corresponding values of $s$, $(\theta_{iz})_1$ and $(\theta_{iz})_0$ are the probabilities that the label $z$ can and cannot be assigned to the sample $x_i$, respectively. 

 \item Compute the empirical marginal distribution vector (of size $2$) with respect to the source label $z$:
 \begin{equation}
 \begin{aligned}
 \hat{P}_z(s) &= \frac{1}{n} \sum_{i=1}^{n} (\theta_{iz})_s \\
 &= \hat{P}_{yz}(0, s) + \hat{P}_{yz}(1, s)
 \end{aligned}
 \end{equation}
 
 \item Compute the $2 \times 2$ empirical conditional distribution matrix $\mathbf{\hat{P}_{y|z}}(t,s)$ of the target label $y$ given the source label $z$, with value at row $t$ column $s$ is: 
 \begin{equation}
 \hat{P}_{y|z}(t|s) = \frac{\hat{P}_{yz}(t, s)}{\hat{P}_z(s)} 
 \end{equation} 
 \end{enumerate} 


For any input $x_i$, consider a binary classifier that predicts whether $x_i$ belongs to label $y$ by first randomly drawing $\mathcal{Z}$ dummy labels from $\Theta(x_i)$, then averaging the likelihood of $y$ based on $\mathcal{Z}$ empirical conditional distributions $\mathbf{\hat{P}_{y|z}}$.
This process is repeated for all $\mathcal{Y}$ target labels.
The set of binary classifiers is called the \textit{\textbf{E}mpirical \textbf{P}redictor} (EP).
MELEP is defined as the average negative log-likelihood of the EP across all target labels, as follows:

 \item Step 3: Compute the Expected Logarithm of Empirical Prediction with respect to the label pair $(y, z)$:

 \begin{equation}
 \label{core_melep}
 \phi (\Theta, \mathcal{D}, y, z) = - \frac{1}{n}\sum_{i=1}^n \log \left(\sum_{s=0}^1 \hat{P}_{y|z}(y_{iz}|s) (\theta_{iz})_s\right)
 \end{equation}
 
 \item Step 4: Compute MELEP by taking the weighted average of $\phi (\theta, \mathcal{D}, y, z)$ over all target-source label pairs:
 \begin{equation}
 \label{eq_melep}
 \Phi (\Theta, \mathcal{D}) = \frac{1}{\mathcal{Y}} 
 \sum_{y} w_y \times \frac{1}{\mathcal{Z}} \sum_{z} \phi (\Theta, \mathcal{D}, y, z)
 \end{equation}

where $w_y$ are the weights of the target label $y$ in the target dataset, i.e. the ratio of the number of positive samples to the number of negative samples of $y$. Note that we do not take the source weights into consideration, because in practice, it makes sense to assume that we do not know the source label distribution prior to fine-tuning.

\end{enumerate}

From its definition, MELEP is always positive, and smaller values indicate superior transferability. Intuitively, MELEP can be regarded as a distance metric, indicating how "close" the pre-trained model $\Theta$ and the target dataset $\mathcal{D}$ are.
The closer the distance, the better the transfer.

The measure is \textit{generic}, meaning that it can be applied to all types of checkpoints, and works without any prior knowledge of the pre-training process, such as data distribution, hyperparameter settings, optimizer, loss functions, etc.
Furthermore, the computation of MELEP is \textit{efficient}, which renders it practically useful. 
This lightweight property is inherited from the original LEEP \cite{leep},
with the calculation involving only a single forward pass through the target dataset $\mathcal{D}$, requiring no training on the downstream task.

\subsection{Datasets}
\label{subsec:datasets}

We used publicly available 12-lead ECG datasets in this work. 
The first source was the public training dataset from the China Physiological Signal Challenge 2018 (CPSC2018) \cite{cpsc2018}. This dataset comprises 6,877 ECG records, each associated with at most nine diagnostic categories: NORM (representing normal ECG patterns), AF (Atrial Fibrillation), I-AVB (First-degree atrioventricular block), LBBB (Left Bundle Branch Block), RBBB (Right Bundle Branch Block), PAC (Premature Atrial Contraction), PVC (Premature ventricular contraction), STD (ST-segment Depression), and STE (ST-segment Elevated).

The second dataset was PTB-XL \cite{ptbxl_dataset}, containing 21,837 records from 18885 patients, and a total of 44 diagnostics statements. 
The dataset's authors organized these diagnostic labels into a hierarchical structure \cite{ptbxl_analysis}, categorizing the 44 labels into five broader superclasses, namely: NORM (normal ECG), MI (Myocardial Infarction), STTC (ST/T-Changes), HYP (Hypertrophy), and CD (Conduction Disturbance). 
We followed this structure and focused on these five superclasses when conducting experiments with the PTB-XL dataset.

Our third dataset, known as the Georgia dataset \cite{georgiadataset}, consists of 10,344 ECGs that reflect the demographic characteristics of the Southeastern United States. 
The data covers a diverse range of 67 unique diagnoses.
However, for our research, we focused on a subset of 10 specific labels, which had the most substantial number of samples: 
NORM, AF, I-AVB, PAC, SB (Sinus Bradycardia), LAD (left axis deviation), STach (Sinus Tachycardia), TAb (T-wave Abnormal), TInv (T-wave Inversion), and LQT (Prolonged QT interval).

The last source was the Chapman University, Shaoxing People's Hospital, and Ningbo First Hospital database \cite{chapman-shaoxing,ningbodata}, which we will refer to as the CSN dataset for brevity. 
This dataset contains 45,152 12-lead ECG records, each lasting for 10 seconds and sampled at 500 Hz.
There are a total of 94 unique labels,
among which we focused on 20 labels with more than 1,000 records for our experiments.
These 20 labels are SB, NORM, STach, TAb, TInv, AF, STD, LAD, PAC, I-AVB, PVC, AFL (Atrial Flutter), LVH (Left Ventricular Hypertrophy), STC (S-T changes), SA (Sinus Arrhythmia), LQRSV (Low QRS Voltages), PR (pacing rhythm), NSTTA (Nonspecific ST-T Abnormality), CRBBB (complete Right Bundle Branch Block), QAb (Q-wave Abnormal).

Table \ref{table:dataset_summary} summarizes key statistics of the four data sources. 
In terms of data preprocessing, we applied the following procedures:

\begin{itemize}
 \item Downsampling: we reduced the sampling frequency of all ECG records from 500 Hz to 100 Hz. This helps reduce computational load while retaining essential information.
 \item Cropping: for ECG records longer than the desired duration (ten seconds), we cropped them to meet this target by keeping only the first ten-second data points. This step ensures that all records have consistent lengths for training.
\end{itemize}

\begin{table*}[ht]
 \centering
 \caption{Statistics of datasets used in this work.}
 \label{table:dataset_summary}
 \begin{tabular}{rcccc}
 \hline
 Dataset & Number of records & Number of labels & Sampling rate (Hz) & Duration (sec) \\
 \hline
 CPSC2018 \cite{cpsc2018} & 6,877 & 9 & 500 & 6-60 \\
 PTB-XL \cite{ptbxl_dataset} & 21,837 & 44 & 500 \& 1000 & 10 \\
 Georgia \cite{georgiadataset} & 10,344 & 62 & 500 & 10 \\
 CSN \cite{chapman-shaoxing,ningbodata} & 45,152 & 94 & 500 & 10 \\
 \hline
 \end{tabular}
 \end{table*}

It is worth noting that only a tiny fraction of records have durations shorter than ten seconds: six out of 6,877 in the CPSC2018 dataset, 52 out of 10,334 in the Georgia dataset, and none in the PTB-XL and CSN datasets.
Therefore, instead of padding these records to meet the desired duration, which could potentially introduce unwanted noise or artifacts into the signals, they were simply omitted from our experiments.

We used the CSN and PTB-XL datasets for fine-tuning due to their relatively large amount of records.
When fine-tuning models on the former, we pre-trained our models using three source datasets: CPSC2018, PTB-XL, and Georgia.
When fine-tuning on the latter, we only used two source datasets: CPSC2018 and Georgia.

For pre-training, we partitioned each of the source datasets into training and test sets as follows.
For PTB-XL, we followed the recommended split in \cite{ptbxl_dataset}, pre-training our models on the first eight folds, and testing on the tenth fold.
For the CPSC2018 and Georgia datasets, we kept 33\% of the amount of data in the test set and allocated the remaining for pre-training.

\subsection{Deep Learning Models}

\begin{figure*}[ht]
 \centering
 \includegraphics[width=\linewidth]{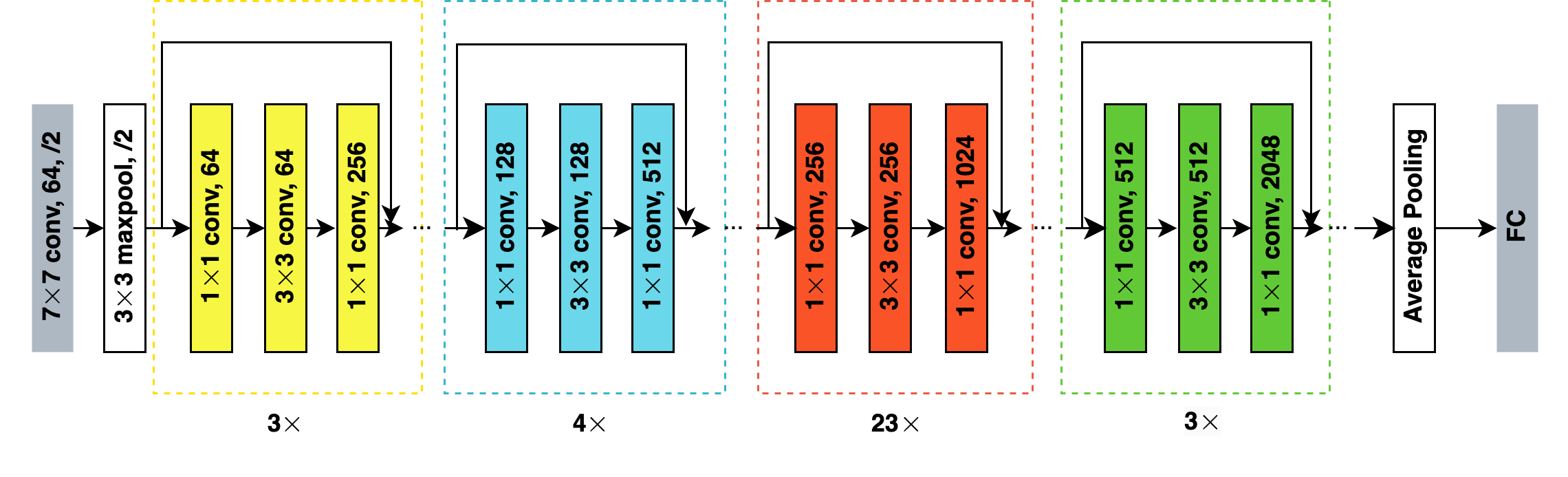}
 \caption{ResNet1d101 Architecture}
 \label{fig:resnet101_arch}
 \end{figure*}

\begin{figure}[t]
 \centering
 \includegraphics[width=0.6\linewidth]{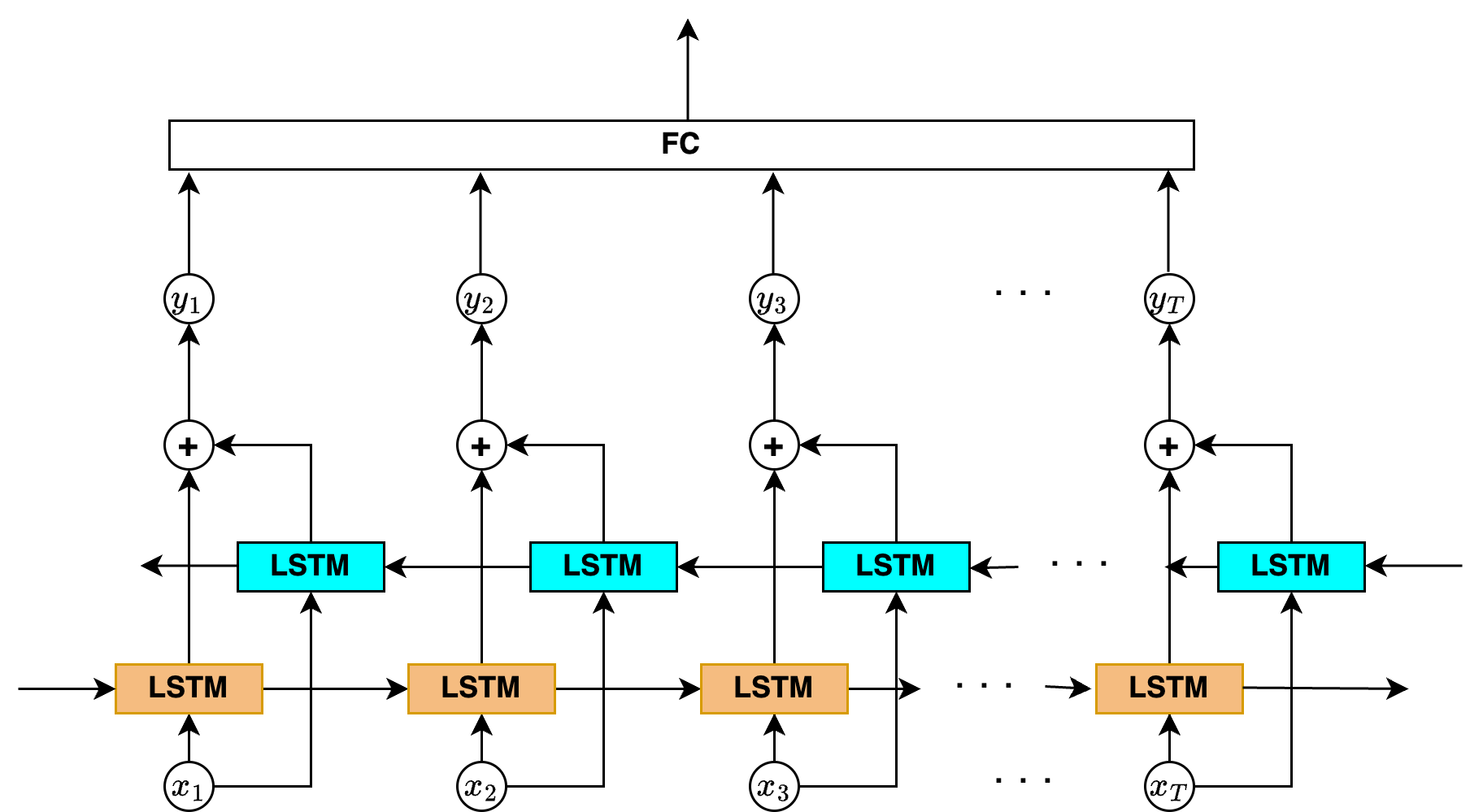}
 \caption{Bi-LSTM Architecture}
 \label{fig:bilstm_arch}
\end{figure}

We investigated two widely used deep learning architectures for time-series analysis:

\begin{itemize}
 \item Convolutional Neural Network (CNN): we utilized ResNet1d101, which is a 1D variant of ResNet101 \cite{resnet101}. The architecture of the ResNet1d101 model is illustrated in Figure \ref{fig:resnet101_arch}. The ResNet family was originally introduced to work with image data, performing well in healthcare applications such as medical imaging \cite{xu2023resnet, harrison2023state, yu2021convolutional, resnetfed}. Yet their power of capturing useful patterns in data still demonstrates strong performance when applied to time-series ECG data \cite{wang201912, zhu2019multi}.
 
 \item Recurrent Neural Network (RNN): the Bidirectional Long Short Term Memory (Bi-LSTM) architecture \cite{lstm} was used. The structure of the Bi-LSTM model is visually presented in Figure \ref{fig:bilstm_arch}. LSTM is also a popular choice when dealing with ECG data, as it is capable of capturing long-term dependencies within the sequences \cite{luo2019multi, mostayed2018classification, lv2019multi, gupta2020transfer}.
\end{itemize}

Since the source datasets have varying numbers of labels, the last fully-connected layer of the models was adjusted to align with the respective number of outputs.
During pre-training, each model was trained on a source training set for 50 epochs, using Adam optimizer \cite{adam} with a learning rate of $0.01$. At the end of each epoch, we recorded the average F1 score on the test set, which served as an early stopping criterion.
We observed that Bi-LSTM experienced overfitting when training beyond the early stopping point, whereas ResNet1d101 mostly converged.

\subsection{Performance metric}

To evaluate the fine-tuning performance of a pre-trained model, we relied on the weighted average F1 score across all labels in the target dataset.
F1 score was chosen as the evaluation metric due to its robustness in handling class imbalances \cite{imbalanced}, a common feature of ECG data, compared to accuracy.

Suppose that each ECG record in the target dataset has $N$ diagnostic labels. Let denote:

\begin{itemize}
 \item $TP_i$: True Positives for the $i^{th}$ label
 \item $FP_i$: False Positives for the $i^{th}$ label
 \item $FN_i$: False Negatives for the $i^{th}$ label
\end{itemize}

Then, the precision ($P_i$), recall ($R_i$), and F1 score ($F_{1_i}$) for the $i^{th}$ label are computed as follows:

\[
P_i = \frac{TP_i}{TP_i + FP_i}
\]

\[
R_i = \frac{TP_i}{TP_i + FN_i}
\]

\[
F_{1i} = \frac{2 \times P_i \times R_i}{P_i + R_i}
\]

The performance metric is computed as the weighted average F1 score across all labels within the target dataset, given by:

\[
 \text{Average F1} = \sum_{i=1}^{N} w_i \times F_{1i}
\]

where the weight $w_i$ represents the ratio of true instances of the $i^{th}$ label.

\section{Experiments \& Results}
\label{sec:results}

In this section, we show the potential of MELEP in predicting the performance of fine-tuning a pre-trained model on a target dataset. 
In practice, transfer learning is often used when dealing with limited human-annotated data.
Therefore, we focused on investigating MELEP in the context of small target datasets.
The code and resources used for experiments can be found at \href{https://github.com/cuongvng/melep-ecg}{github.com/cuongvng/melep-ecg}.

\subsection{Relation between MELEP and Model Performance of CNN fine-tuned on CSN}
\label{exp_resnet_shaoxing}

We first experimented with the convolutional model ResNet1d101.
This model was pre-trained on three different source datasets: PTB-XL, CPSC2018, and Georgia, as described in Section \ref{subsec:datasets}, resulting in three respective source checkpoints. 

Each source checkpoint was then undergone an experiment with a wide range of target tasks sampled from the CSN dataset.
To construct these tasks, we started with the set of 20 labels in the CSN dataset with at least 1000 positive samples, as in Section \ref{subsec:datasets}.
$N$ labels were then randomly sampled without replacement from the set, where $N$ varied from 2 to 10.
This step ensured that the target tasks would cover a diverse set of target labels.
Records with no positive values for the $N$ selected labels were filtered out to avoid creating a sparse dataset, and to guarantee that every sample left contained at least one positive label.
We then randomly select 1000 records among the remaining to form a data fold.
The process was repeated 100 times to generate a total of 100 data folds for our experiment.

\begin{figure}[t]

 \begin{minipage}[t]{0.3\linewidth}
 \centering
 \centerline{\includegraphics[width=4cm]{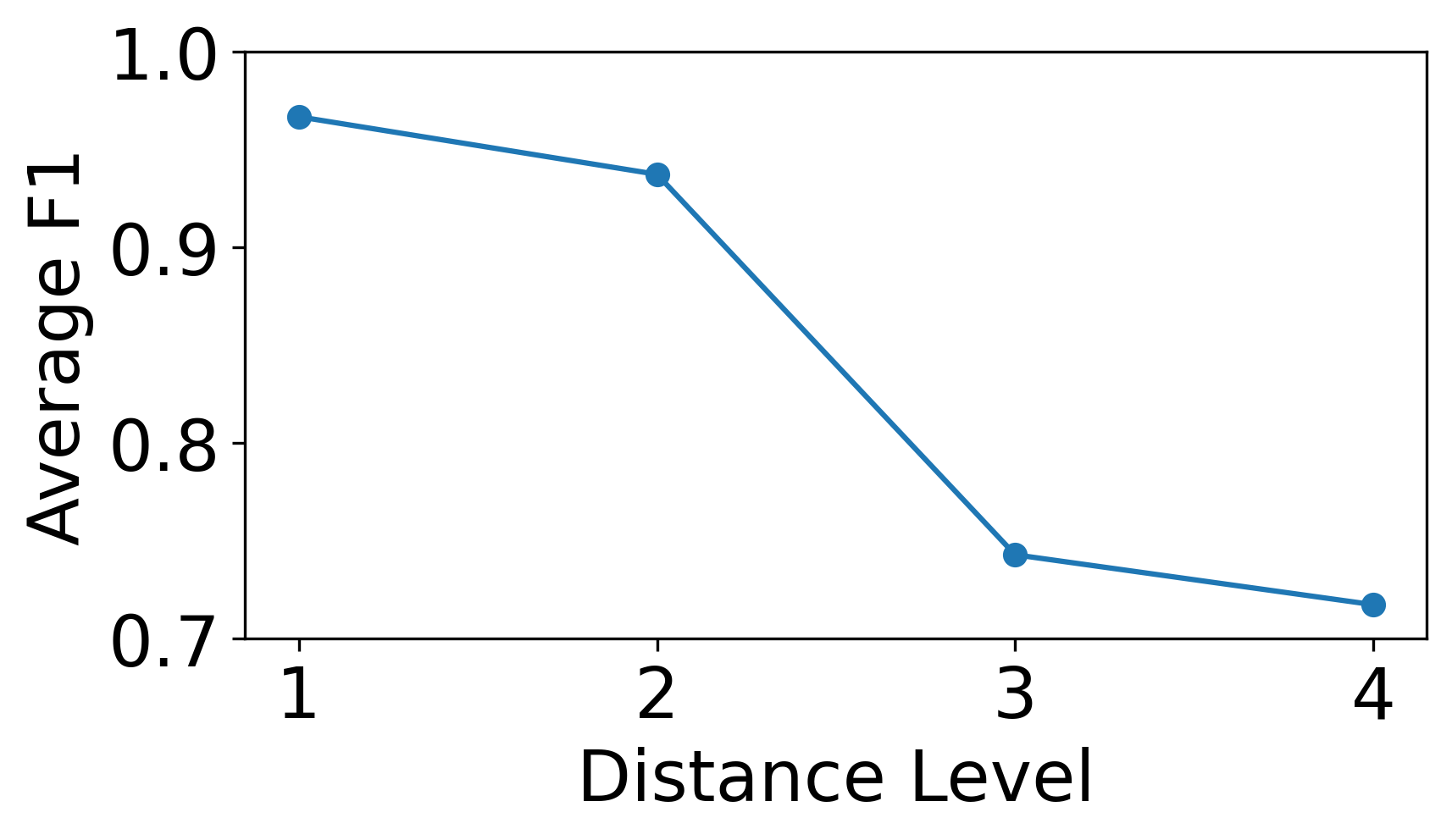}}
 \centerline{\footnotesize (a) PTB-XL $\rightarrow$ CSN}\medskip
 \end{minipage}
 \hfill
 \begin{minipage}[t]{.3\linewidth}
 \centering
 \centerline{\includegraphics[width=4.0cm]{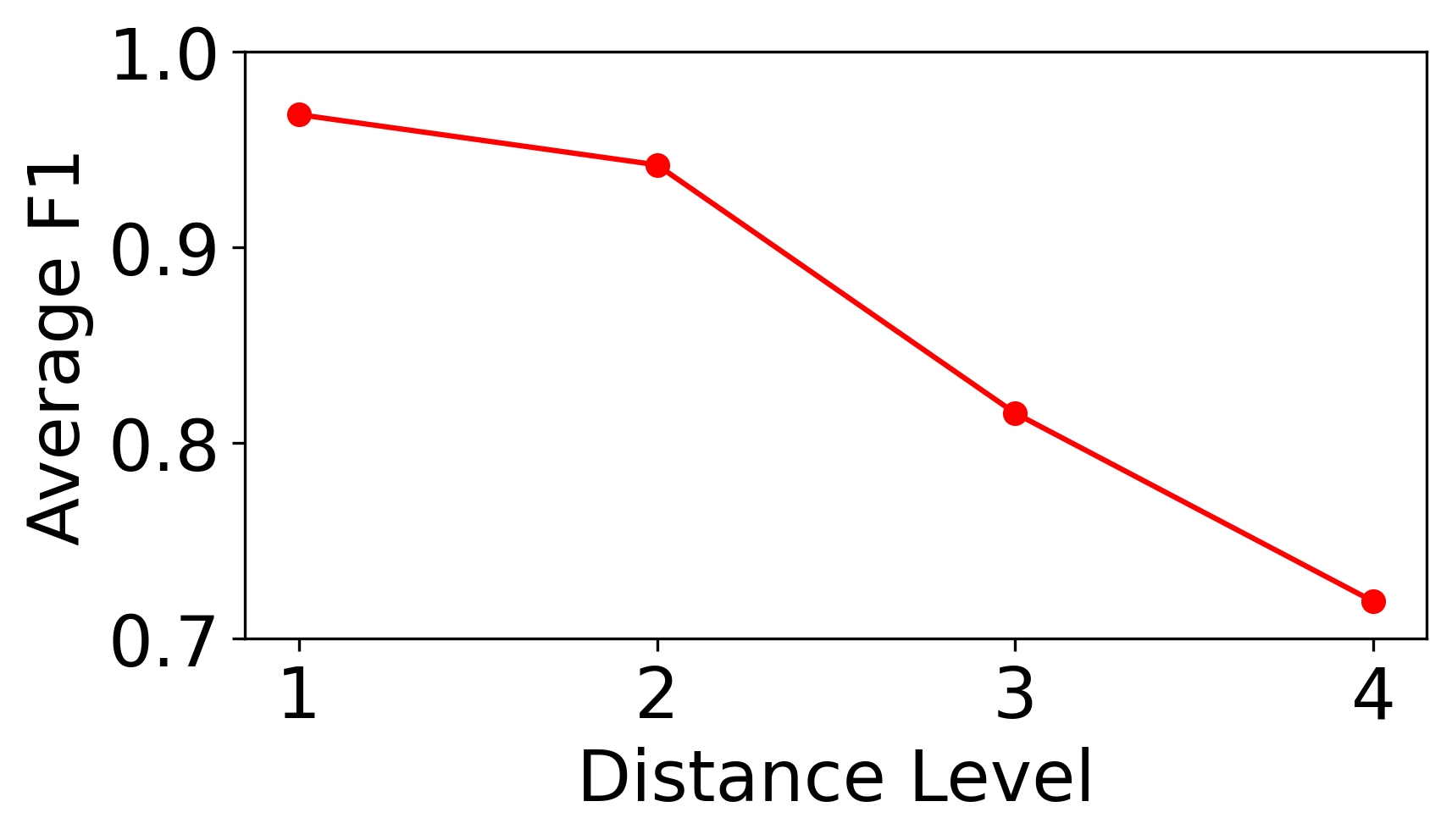}}
 \centerline{\footnotesize (b) CPSC2018 $\rightarrow$ CSN}\medskip
 \end{minipage}
 \hfill
 \begin{minipage}[t]{0.3\linewidth}
 \centering
 \centerline{\includegraphics[width=4.0cm]{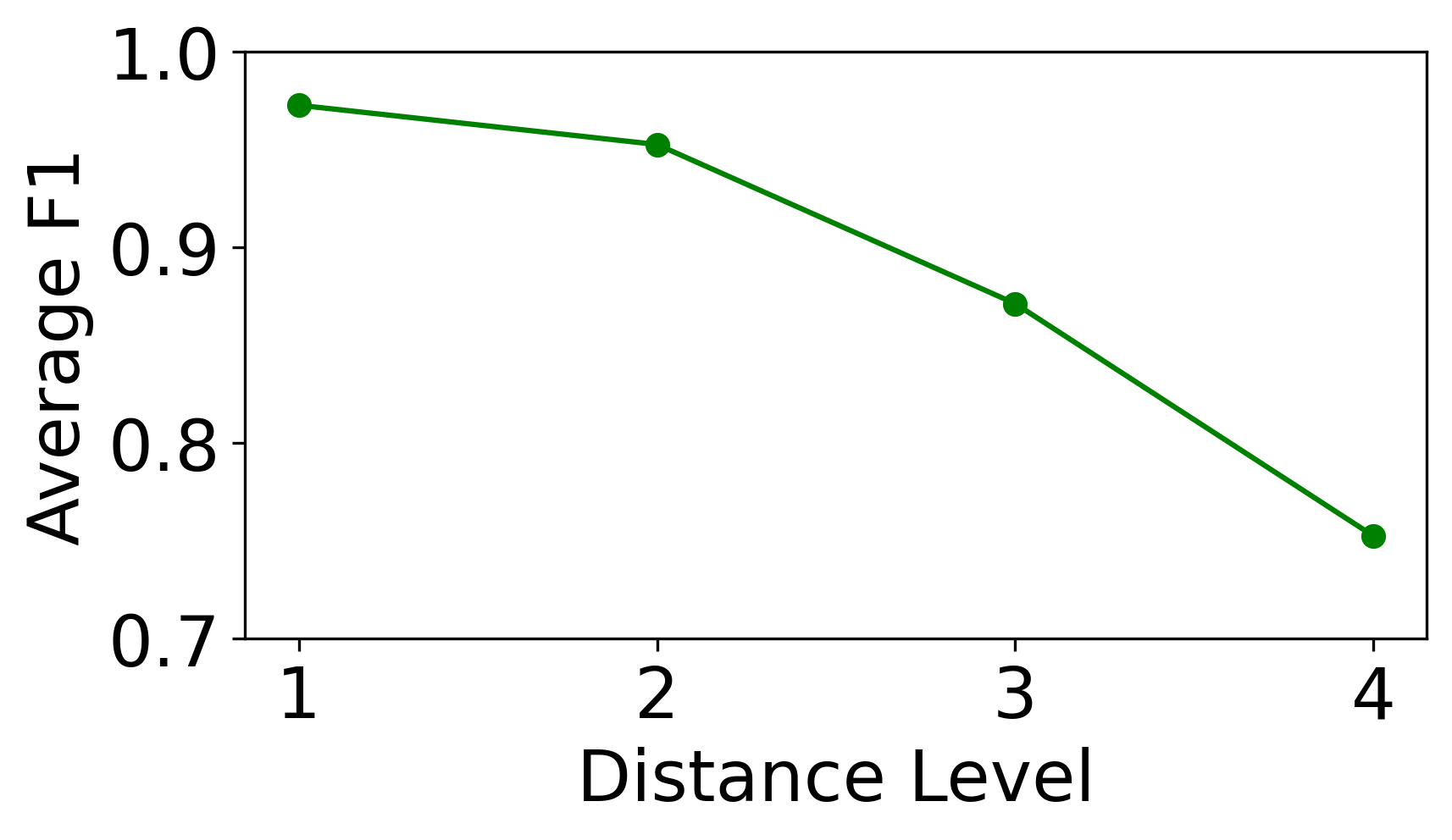}}
 \centerline{ \footnotesize (c) Georgia $\rightarrow$ CSN}\medskip
 \end{minipage}
 \caption{Relation of MELEP (partitioned as four distance levels) and fine-tuning performance of ResNet1d101 on target tasks sampled from the CSN dataset. The lower the MELEP (the closer the distance), the better transferability.}
 \label{fig:resnet101_shaoxing}
\end{figure}

For each fold, we further split it into training and test subsets with a 7:3 ratio, i.e. 700 training records and 300 test records.
Subsequently, we compute MELEP using the pre-trained checkpoint and the training subset only, following the algorithm described in Section \ref{subsec:melep}.
Prior to fine-tuning the model, we replaced the top fully connected layer of the checkpoint, adjusting the number of output neurons to match the target number of labels $N$.
The entire modified model was then fine-tuned on the training subset for 50 epochs with early stopping, using Adam optimizer \cite{adam}, and then evaluated on the test subset using weighted average F1 score across the $N$ labels of the given fold. 
Ultimately, we gathered 100 points of (MELEP, average F1) representing the correlation between MELEP and the fine-tuning performance of the source checkpoint across a wide range of target tasks.

We then performed a Pearson correlation analysis between MELEP and the performance, following a similar approach used in assessing transferability on multi-class computer vision tasks \cite{tran2019transferability,leep}.
The first three rows in Table \ref{table:full_results} show the results of the three ResNet1d101 checkpoints in this experiment, revealing strong negative correlations between MELEP and average F1 scores, all of which are below $-0.6$.
To visualize this relationship, Figure \ref{fig:resnet101_shaoxing} classifies the MELEP values into four distinct distance levels.
Within each level, we calculated the mean of average F1 scores from all the folds with MELEP falling into that level. The lower the MELEP, the closer the distance, implying easier transferability.

\subsection{Relation between MELEP and Model Performance of RNN fine-tuned on CSN}
\label{exp_bilstm_shaoxing}

To illustrate the applicability of MELEP to RNN, we repeated the experiment in Section \ref{exp_resnet_shaoxing} with Bi-LSTM as the source model. 
Similar to ResNet1d101, the Bi-LSTM model was pre-trained on three source datasets: PTB-XL, CPSC2018, and Georgia.
We leveraged the same set of 100 CSN data folds which were previously constructed for the CNN experiment, and applied the identical fine-tuning procedure.

\begin{figure}[t!]

 \begin{minipage}[t]{.3\linewidth}
 \centering
 \centerline{\includegraphics[width=4cm]{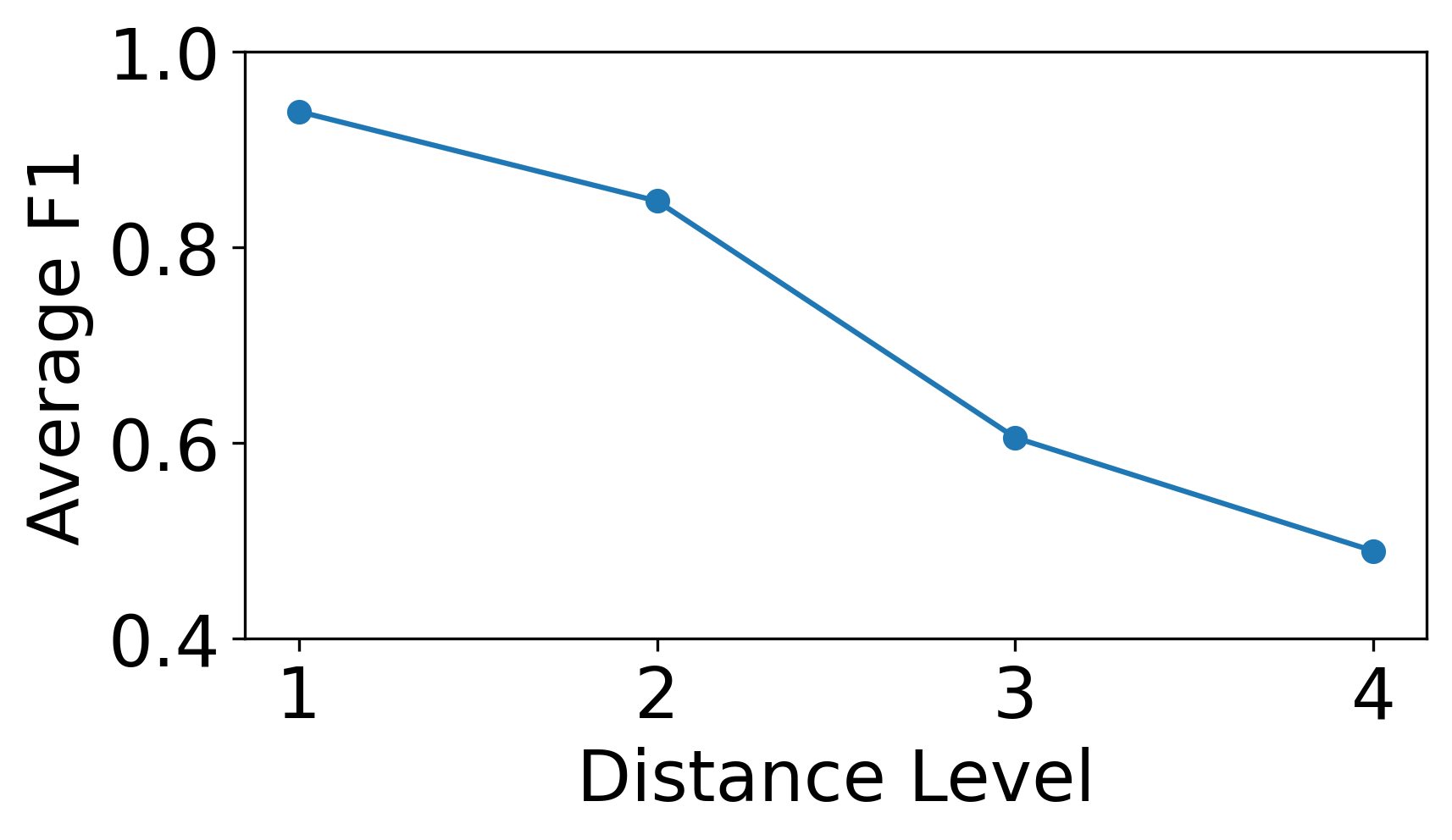}}
 \centerline{\footnotesize (a) PTB-XL $\rightarrow$ CSN}\medskip
 \end{minipage}
 \hfill
 \begin{minipage}[t]{.3\linewidth}
 \centering
 \centerline{\includegraphics[width=4cm]{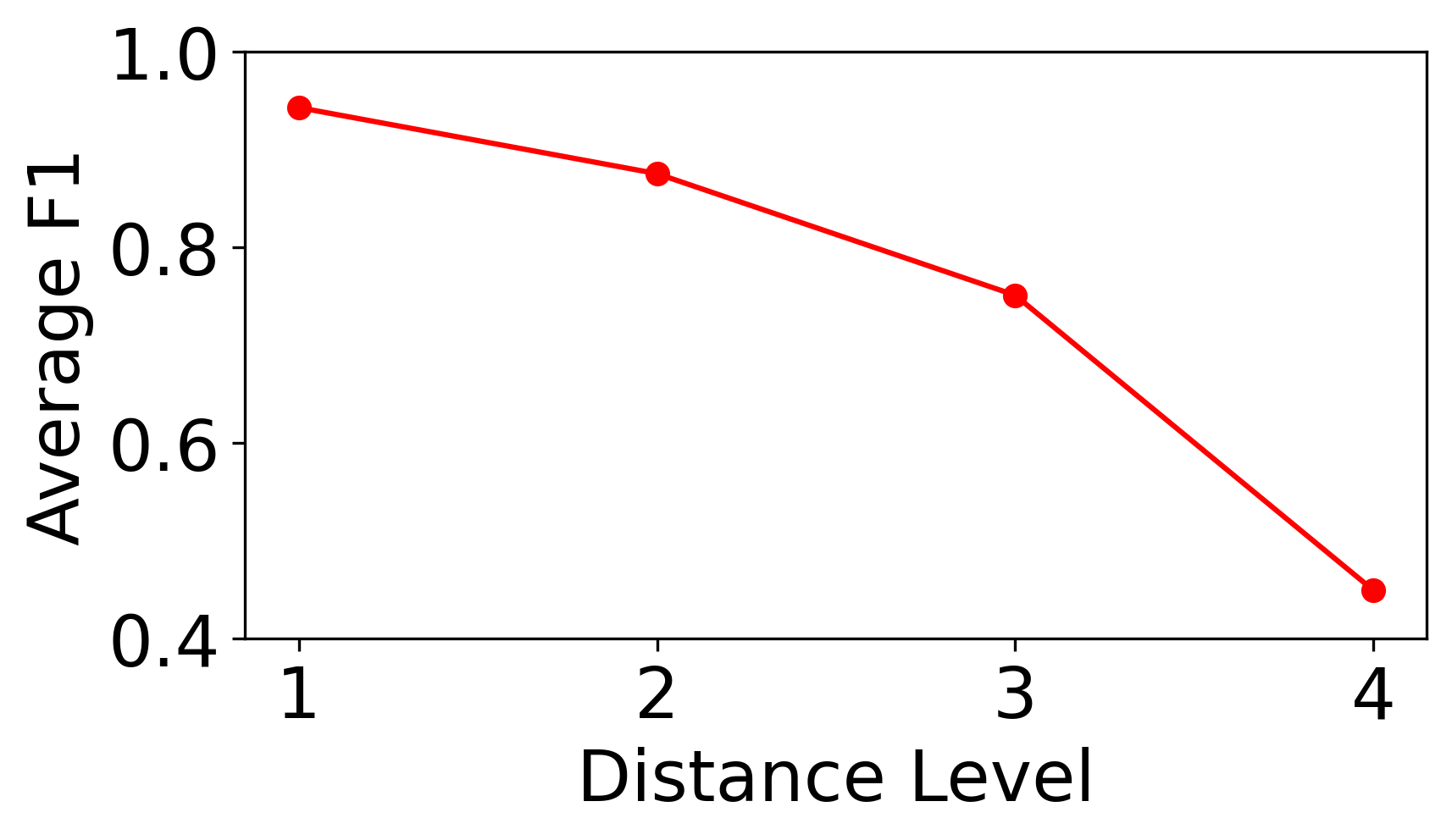}}
 \centerline{\footnotesize (b) CPSC2018 $\rightarrow$ CSN}\medskip
 \end{minipage}
 \hfill
 \begin{minipage}[t]{.3\linewidth}
 \centering
 \centerline{\includegraphics[width=4cm]{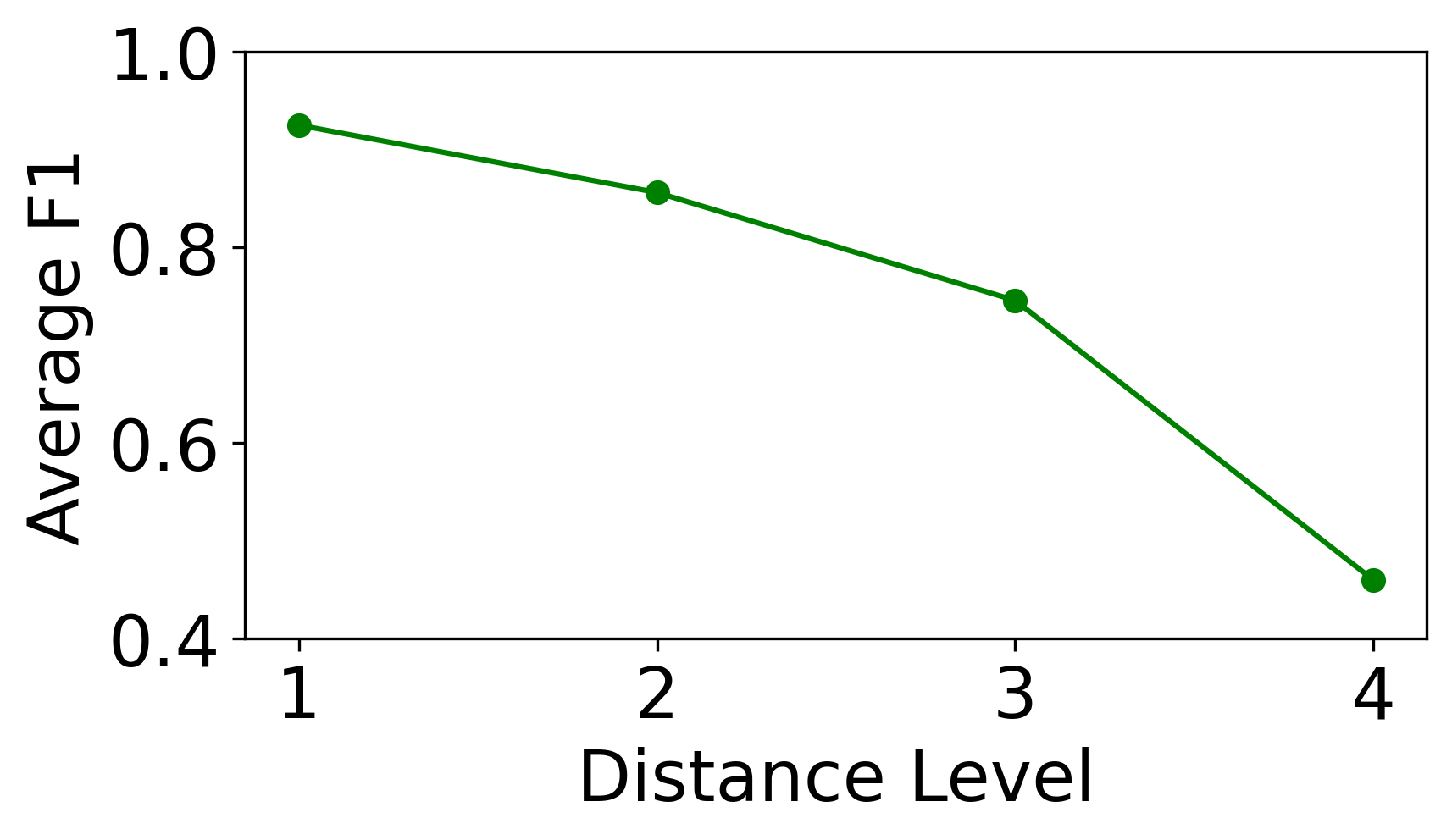}}
 \centerline{\footnotesize (c) Georgia $\rightarrow$ CSN}\medskip
 \end{minipage}
 \caption{Relation of MELEP (partitioned as four distance levels) and fine-tuning performance of Bi-LSTM on target tasks sampled from the CSN dataset. The lower the MELEP (the closer the distance), the better transferability.}
 \label{fig:bilstm_shaoxing}
\end{figure}

In Table \ref{table:full_results} (specifically, the first three rows of the Bi-LSTM section), we observe a robust correlation, even stronger than that observed with ResNet1d101, between MELEP and average F1 scores. This correlation is visually depicted in Figure \ref{fig:bilstm_shaoxing}, where MELEP is categorized into the same distance levels as described in Section \ref{exp_resnet_shaoxing}.
The trend remains consistent: the closer the distance, the better the transfer.

\subsection{Relation between MELEP and Performance of Models fine-tuned on PTB-XL}
\label{exp_ptbxl}

In this experiment, we explored the use of MELEP on a different target dataset, specifically PTB-XL, chosen for its relatively large amount of records.
We followed the same procedure outlined in Section \ref{exp_resnet_shaoxing} to construct 100 target data folds, with the only difference being the number of labels $N$.
These label sets ranged from two to five and were derived from the five superclasses covering the whole PTB-XL dataset, as described in Section \ref{subsec:datasets}. 

\begin{figure}[t]
 \begin{minipage}[t]{0.45\linewidth}
 \centering
 \centerline{\includegraphics[width=4.0cm]{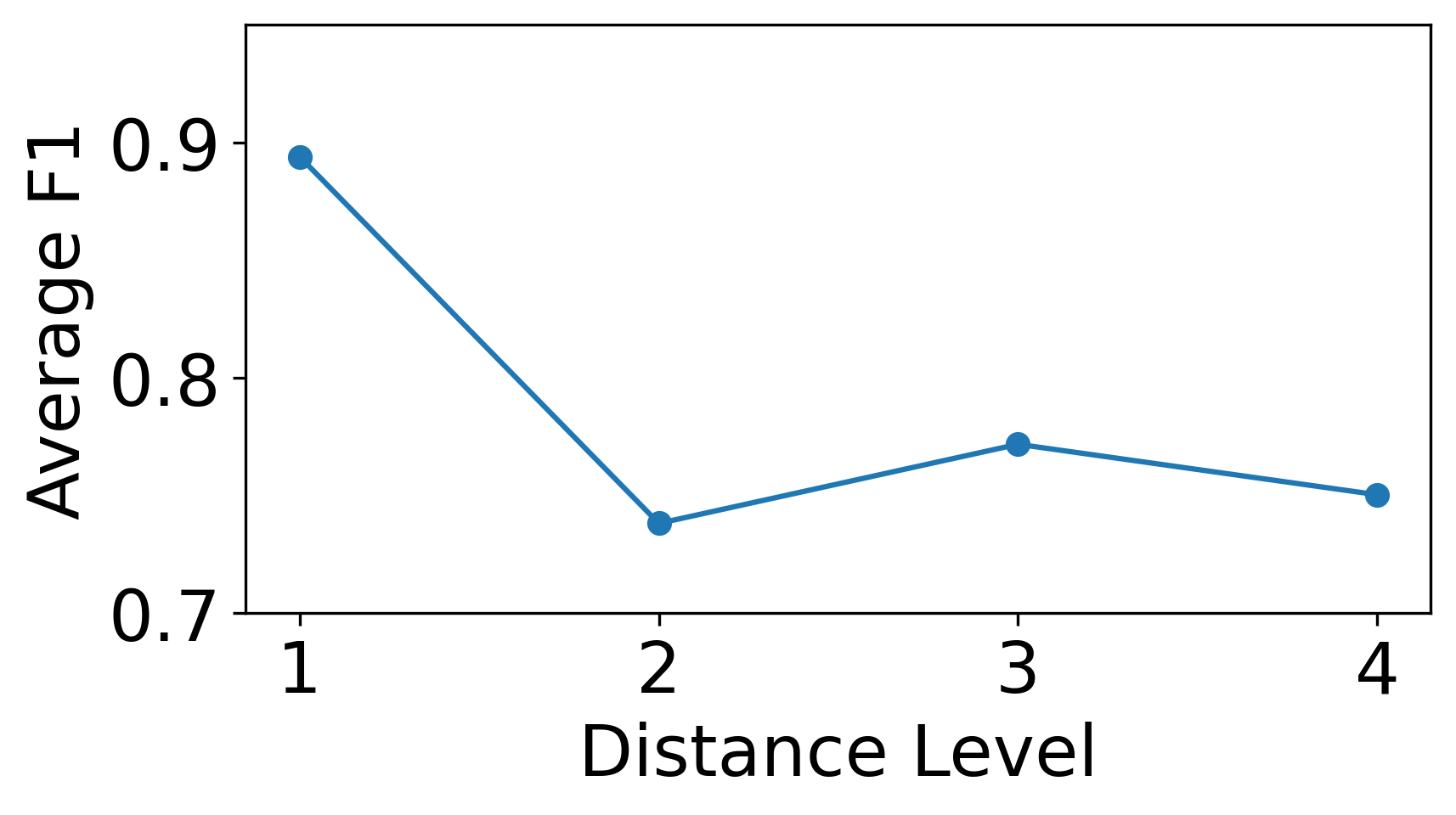}}
 \centerline{\footnotesize (a) ResNet1d101} 
 \centerline{\footnotesize CPSC2018 $\rightarrow$ PTB-XL}\medskip
 \end{minipage}
 \hfill
 \begin{minipage}[t]{0.45\linewidth}
 \centering
 \centerline{\includegraphics[width=4.0cm]{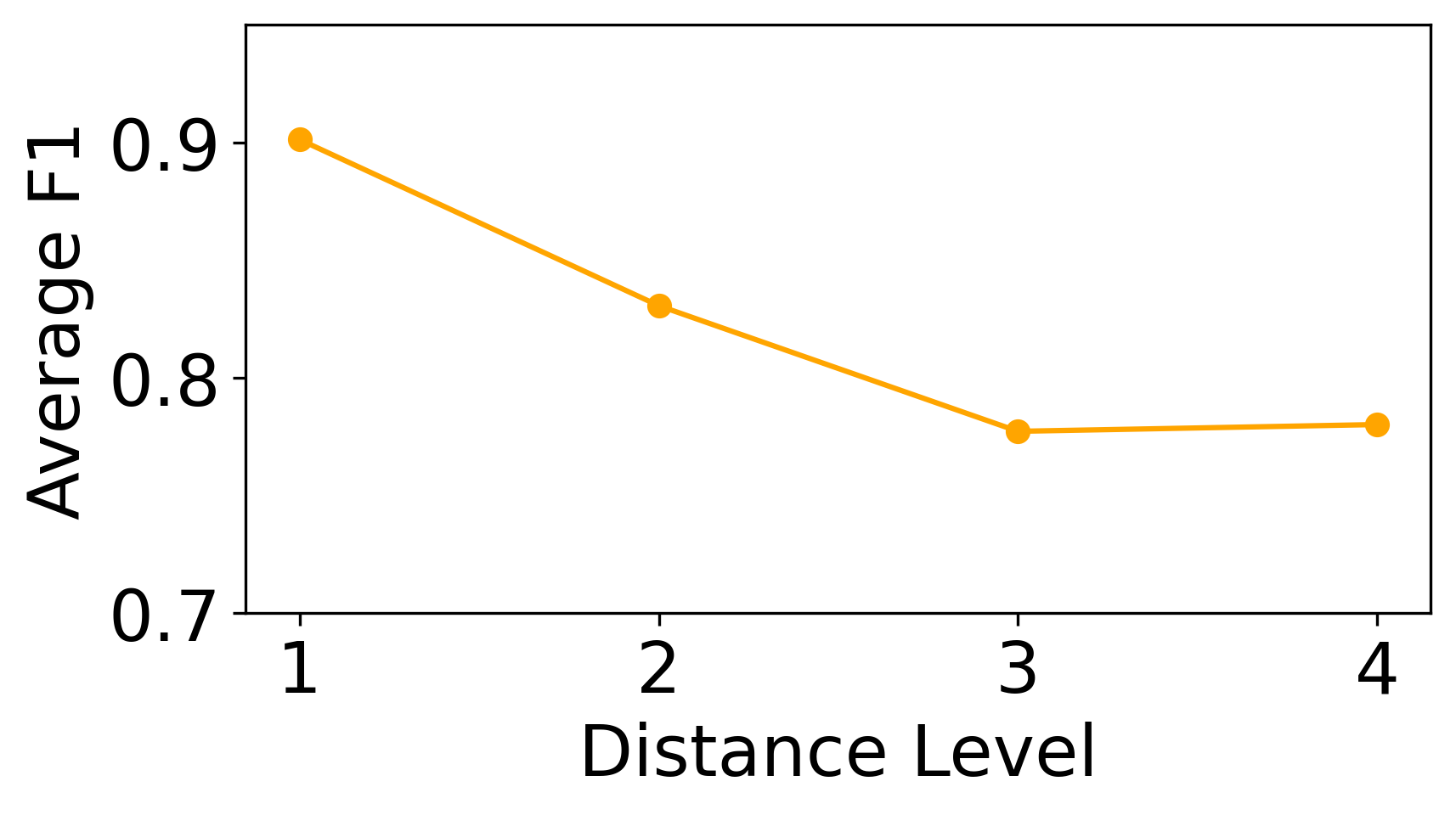}}
 \centerline{\footnotesize (b) ResNet1d101}
 \centerline{\footnotesize Georgia $\rightarrow$ PTB-XL}\medskip
 \end{minipage}
 
 \begin{minipage}[t]{.45\linewidth}
 \centering
 \centerline{\includegraphics[width=4.0cm]{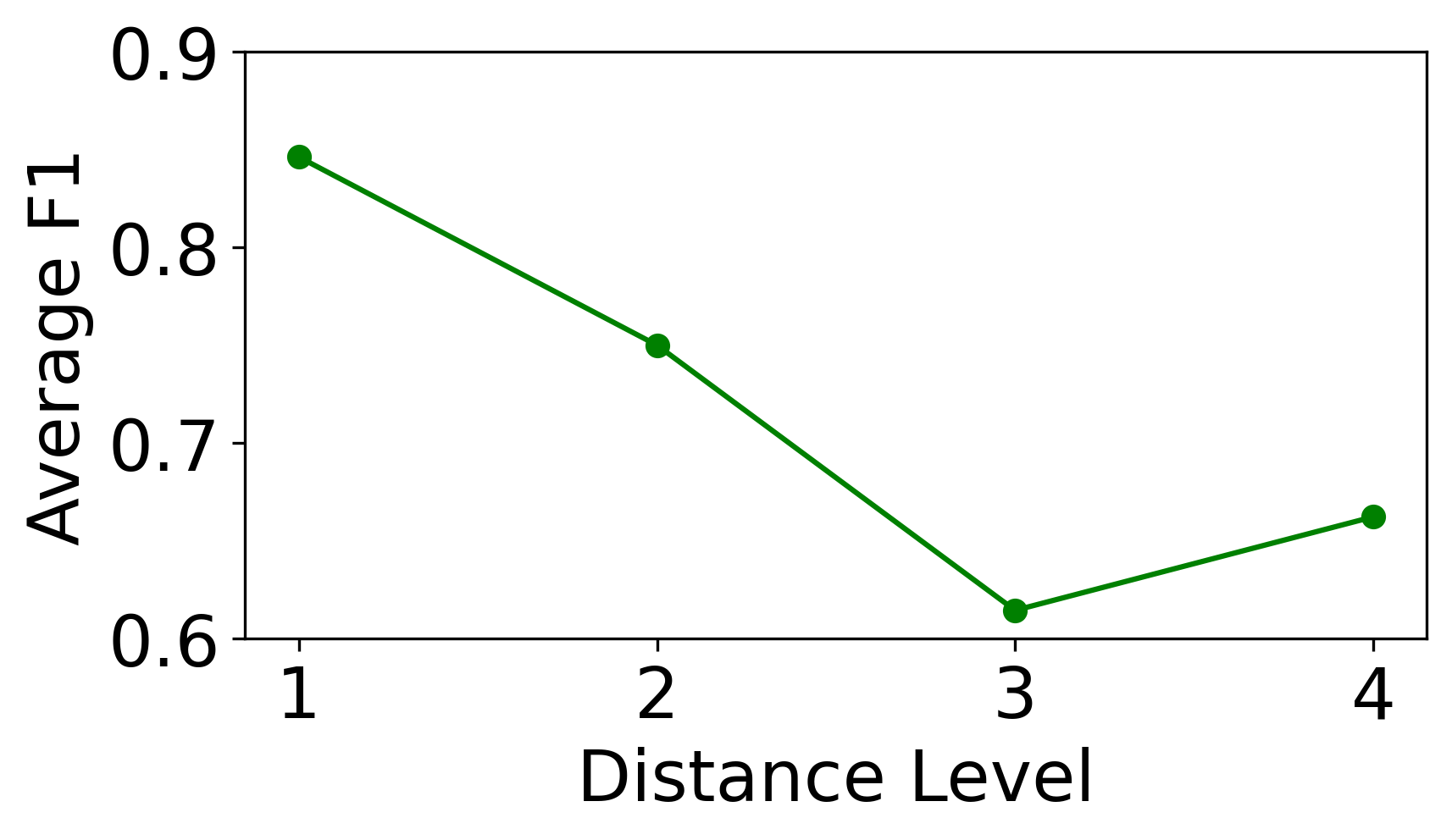}}
 \centerline{\footnotesize (c) Bi-LSTM}
 \centerline{\footnotesize CPSC2018 $\rightarrow$ PTB-XL}\medskip
 \end{minipage}
 \hfill
 \begin{minipage}[t]{0.45\linewidth}
 \centering
 \centerline{\includegraphics[width=4.0cm]{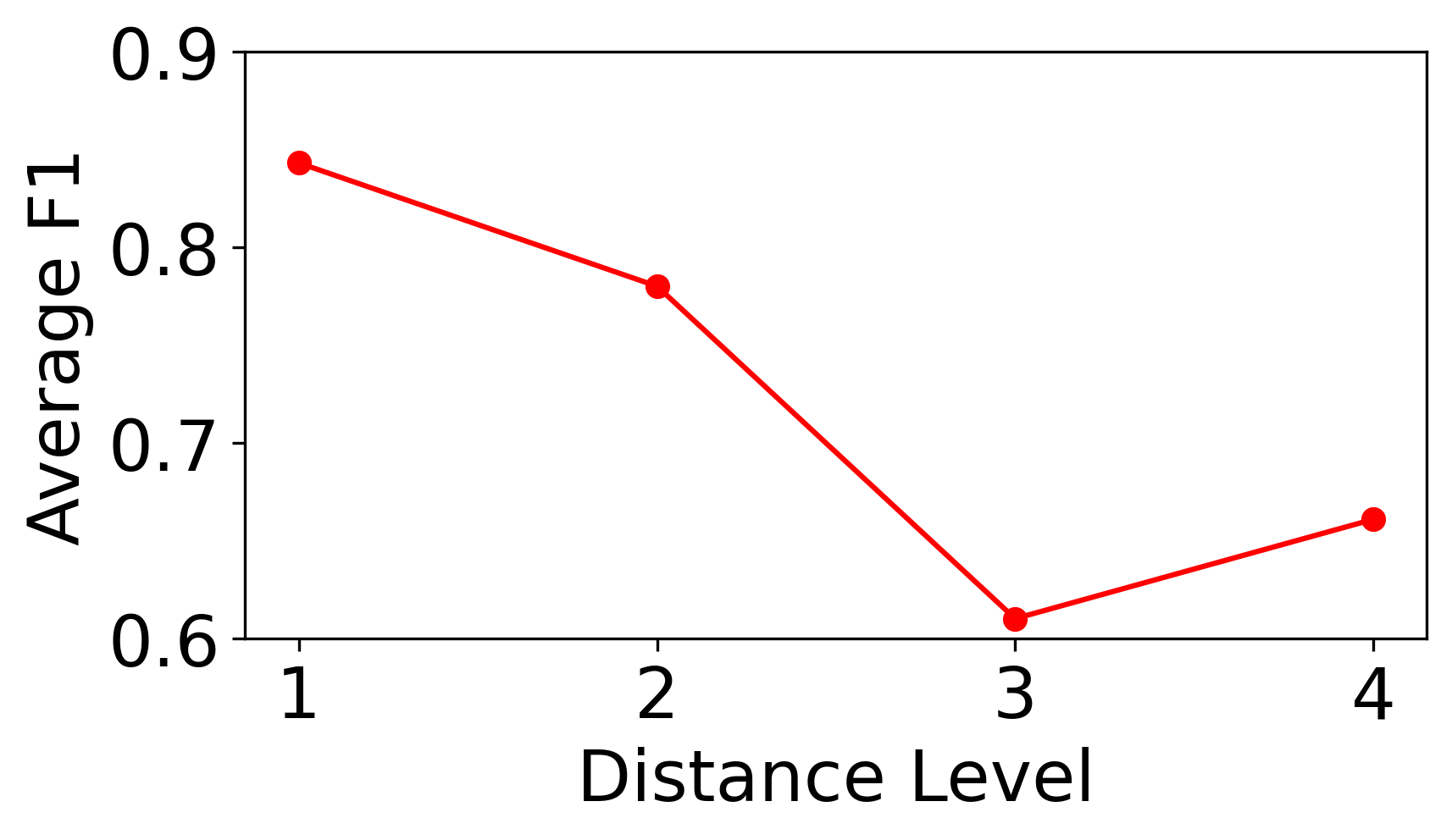}}
 \centerline{\footnotesize (d) Bi-LSTM}
 \centerline{\footnotesize Georgia $\rightarrow$ PTB-XL}\medskip
 \end{minipage}
 \caption{Relation of MELEP (partitioned as four distance levels) and fine-tuning performance of ResNet1d101 and Bi-LSTM on target tasks sampled from the PTB-XL dataset.}
 \label{fig:to_ptbxl}
\end{figure}

We considered four different checkpoints: ResNet1d101 and Bi-LSTM models pre-trained on the CPSC2018 and Georgia datasets. 
The results in Table \ref{table:full_results} indicate a moderate correlation between MELEP and transfer performance, with most correlation coefficients below -0.5.
These correlations, while still significant, are slightly weaker than what was observed in the experiment with the CSN dataset (Section \ref{exp_resnet_shaoxing} and \ref{exp_bilstm_shaoxing}), as shown in Figure \ref{fig:to_ptbxl}, where the predictive trend of MELEP is disrupted, with an increase instead of a decrease at one distance level (the $2^{nd}$ level for ResNet1d101 pre-trained on CPSC2018 and the $3^{rd}$ level for other checkpoints).

\begin{table*}[t]
 \centering
 \def\arraystretch{1.2}%
 \caption{Pearson correlation coefficients between MELEP and average of F1 scores in the experiments described in Section \ref{exp_resnet_shaoxing}, \ref{exp_bilstm_shaoxing} and \ref{exp_ptbxl}. Strong negative correlations were observed for most cases, indicating MELEP's potential to predict fine-tuning performance with only a single forward pass required. All correlations are statistically significant.}
 \label{table:full_results}
 \begin{tabular}{cccccc}
 \hline
 Model & Source Data & Target Data & Details in & \makecell{Pearson \\ Correlation ($r$)} & $p$-value \\
 \hline
 \multirow{5}{*}{ResNet1d101}
 & PTB-XL & CSN & Sec. \ref{exp_resnet_shaoxing} & -0.639 & $8.1 \times 10^{-13}$ \\
 & CPSC2018 & CSN & Sec. \ref{exp_resnet_shaoxing} & -0.631 & $2.0 \times 10^{-12}$\\
 & Georgia & CSN & Sec. \ref{exp_resnet_shaoxing} & -0.608 & $1.9\times10^{-11}$\\
 & CPSC2018 & PTB-XL & Sec. \ref{exp_ptbxl} & -0.476 & $5.7 \times 10^{-7}$ \\
 & Georgia & PTB-XL & Sec. \ref{exp_ptbxl} & -0.500 & $1.1 \times 10^{-7}$ \\
 \hline
 \multirow{5}{*}{Bi-LSTM}
 & PTB-XL & CSN & Sec. \ref{exp_bilstm_shaoxing} & -0.691 & $1.7 \times 10^{-15}$ \\
 & CPSC2018 & CSN & Sec. \ref{exp_bilstm_shaoxing} & -0.670 & $2.6 \times 10^{-14}$ \\
 & Georgia & CSN & Sec. \ref{exp_bilstm_shaoxing} & -0.665 & $4.2 \times 10^{-14}$ \\
 & CPSC2018 & PTB-XL & Sec. \ref{exp_ptbxl} & -0.551 & $2.8 \times 10^{-9}$ \\
 & Georgia & PTB-XL & Sec. \ref{exp_ptbxl} & -0.517 & $3.5 \times 10^{-8}$ \\
 \hline
 \end{tabular}
 \end{table*}

\subsection{MELEP for Checkpoint Selection}
\label{subsec:ckp_selection}
This experiment demonstrates the use of MELEP in practice to effectively estimate fine-tuning performance in a multi-label classification task before the actual fine-tuning process takes place.
Consider a checkpoint selection problem, where the goal is to choose the best candidate from a set of given source checkpoints for a target task. 

In this scenario, we had eight checkpoint candidates: ResNet1d101-PTBXL, ResNet1d101-CPSC, ResNet1d101-Georgia, ResNet1d101-CSN, BiLSTM-PTBXL, BiLSTM-CPSC, BiLSTM-Georgia, BiLSTM-CSN. These checkpoints were obtained by pretraining two DNNs (ResNet1d101 and BiLSTM) on four datasets (PTBXL, CPSC2018, Georgia, and CSN).
To simulate the context of fine-tuning a small target dataset, for each of the four datasets, we generated four target folds of 1000 records, following the random process outlined in Section \ref{exp_resnet_shaoxing}, with a full set of 5, 9, 10 and 20 labels, respectively.
For a given target fold, two checkpoints pre-trained on the same dataset were excluded to ensure fair comparison.
For example, we did not consider the ResNet1d101-PTBXL and BiLSTM-PTBXL for experiments with the target PTBXL fold.
Subsequently, we divided each fold into training and test subsets with a 7:3 ratio. The training subset was used for computing MELEP and fine-tuning, while the test set was reserved for performance evaluation.


\begin{figure}[t!]

 \begin{minipage}[t]{0.45\linewidth}
 \centering
 \centerline{\includegraphics[width=6.0cm]{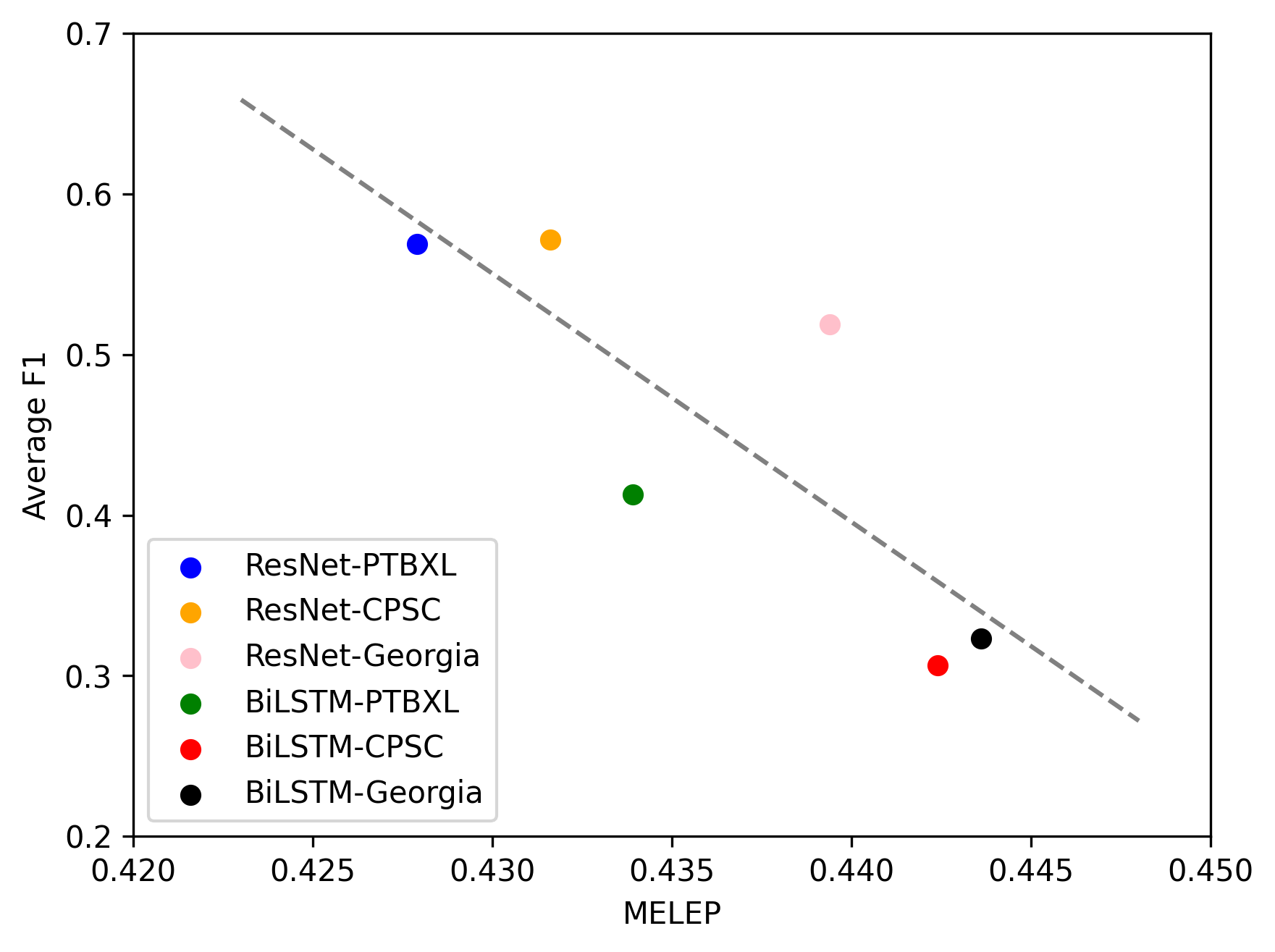}}
 \centerline{\footnotesize (a) CSN} \medskip 
 \end{minipage}
 \hfill
 \begin{minipage}[t]{0.45\linewidth}
 \centering
 \centerline{\includegraphics[width=6.0cm]{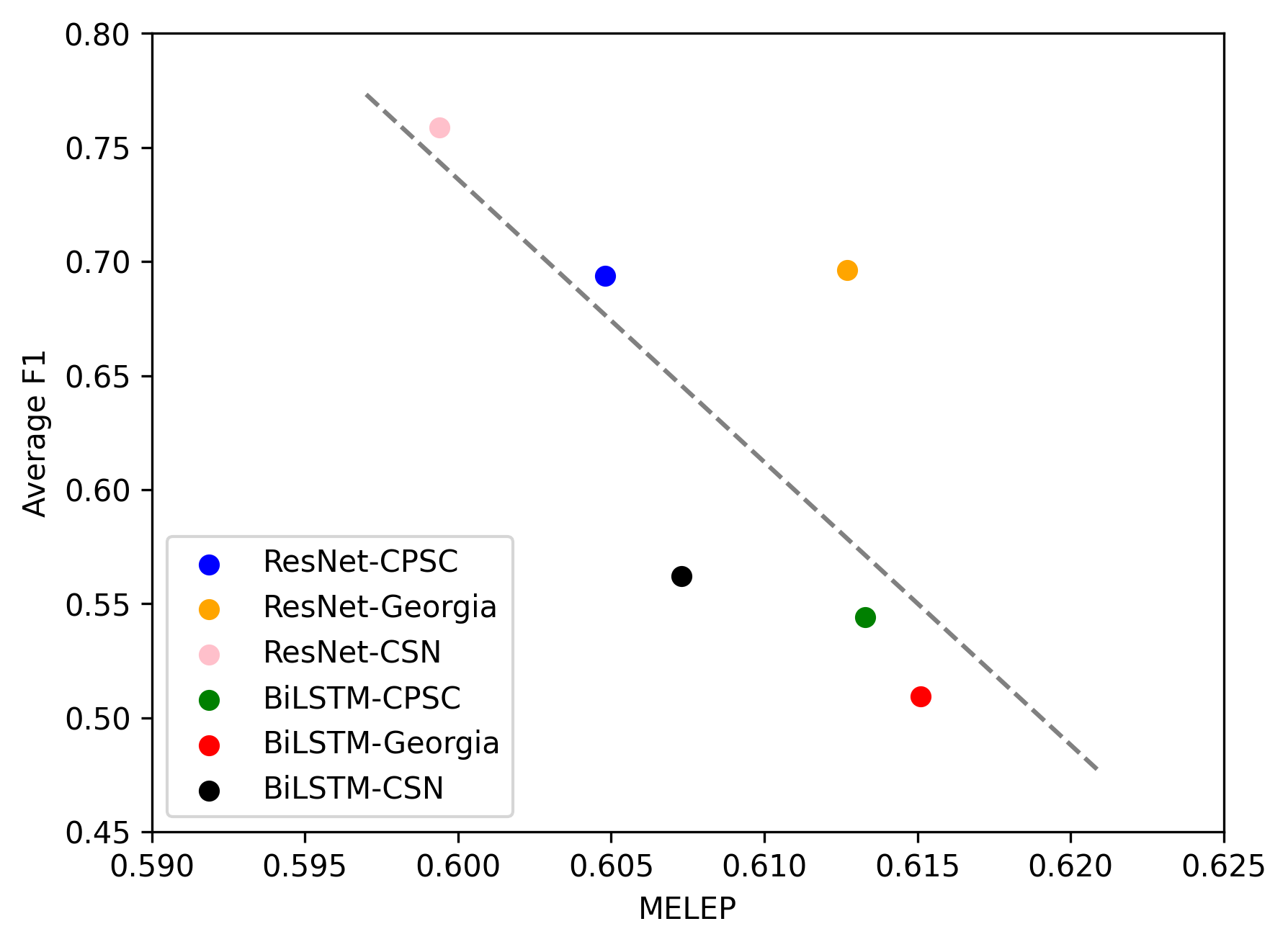}}
 \centerline{\footnotesize (b) PTB-XL}\medskip
 \end{minipage}
 
 \begin{minipage}[t]{.45\linewidth}
 \centering
 \centerline{\includegraphics[width=6.0cm]{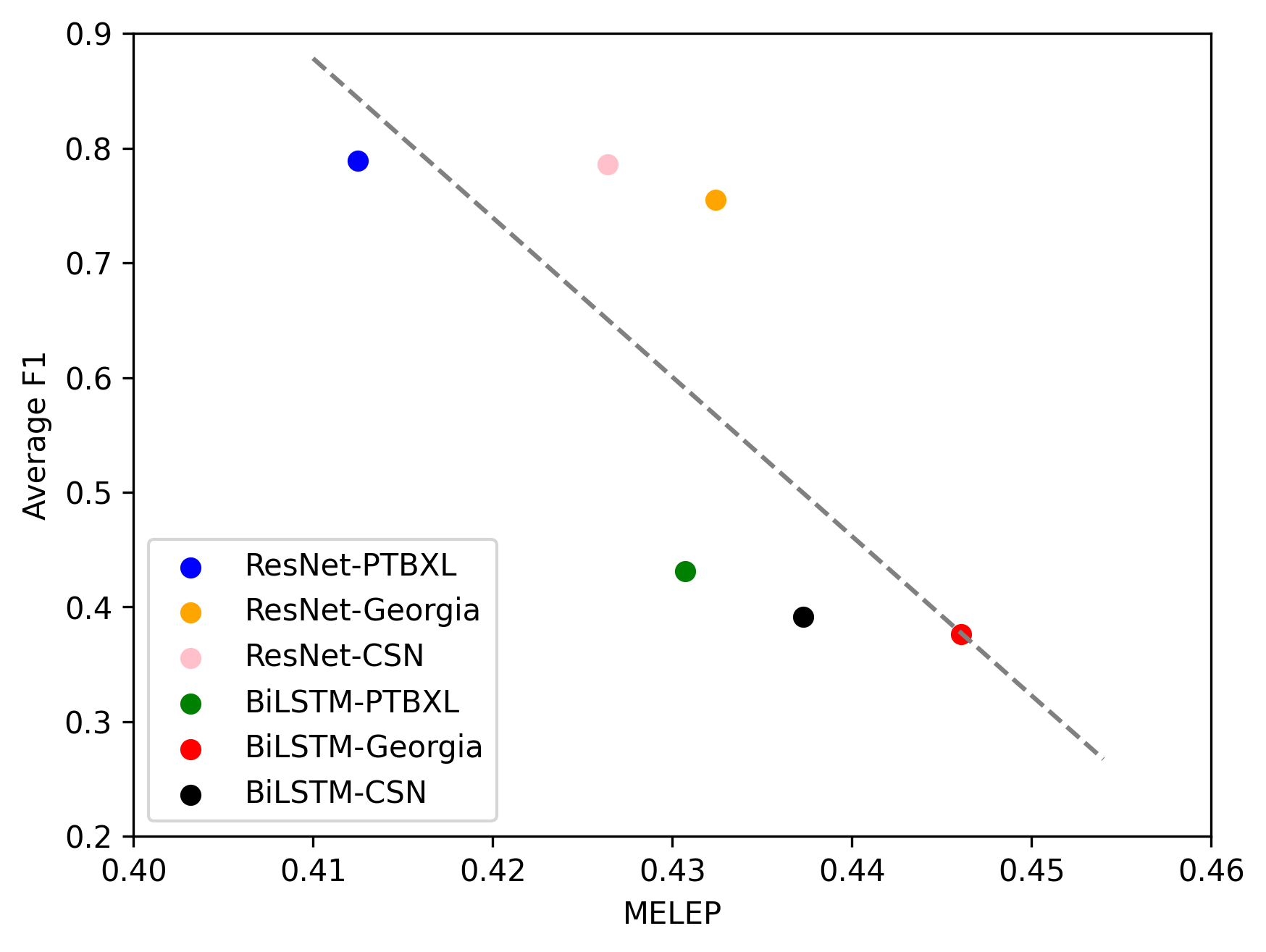}}
 \centerline{\footnotesize (c) CPSC2018}\medskip
 \end{minipage}
 \hfill
 \begin{minipage}[t]{0.45\linewidth}
 \centering
 \centerline{\includegraphics[width=6.0cm]{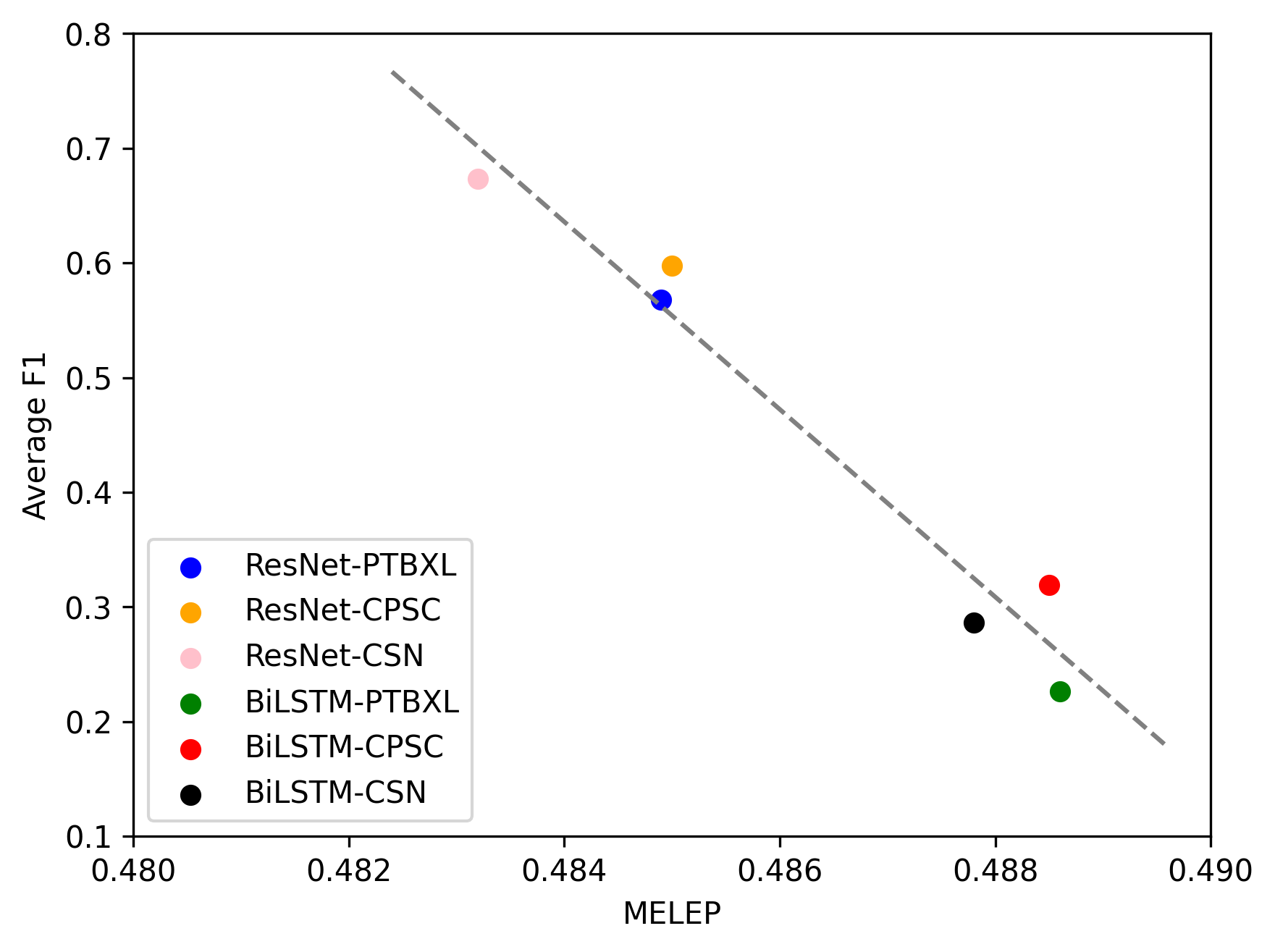}}
 \centerline{\footnotesize (d) Georgia} \medskip
 \end{minipage}
 \caption{MELEP for checkpoint selection problem with corresponding target dataset. The consistent trend demonstrates MELEP's effectiveness in predicting the fine-tuning performance, supporting the pre-selection of the best pre-trained models.}
 \label{fig:checkpoint_selection}
\end{figure}

In Figure \ref{fig:checkpoint_selection}, we display the MELEP values and their corresponding average F1 scores for all checkpoint candidates for each target task, along with the reference best-fit lines. 
The four graphs illustrate the effectiveness of MELEP in predicting the performance of the given checkpoints on the target task: lower MELEP should indicate a better average F1 score.

\section{Discussions \& Conclusion}
\label{sec:discussion}

We introduced MELEP, a novel transferability measure that is directly applicable to multi-label ECG diagnosis.
The measure is built upon the foundation of LEEP \cite{leep}, adapting from single-label multi-class problems in computer vision to multi-label binary-class ones in the ECG domain. 
We conducted extensive experiments to empirically illustrate the effectiveness of MELEP in predicting the performance of transfer learning in various ECG classification tasks.
In this section, we discuss some notable properties, extensions, and applications of MELEP alongside promising directions for future study.

\textbf{Source model dependence}: MELEP computation is based on a source checkpoint, which is a source model pre-trained on a source task.
This may imply that better performance of the source model may result in better MELEP (thus better transferability).
Empirically, in our experiments, pre-trained ResNet outperformed pre-trained Bi-LSTM on all source tasks, indicated in Table \ref{table:pretrained_performance}. 
Finetuned ResNet mostly achieved better MELEP scores than and outperformed fine-tuned Bi-LSTM on most target tasks in the checkpoint selection experiment in Section \ref{subsec:ckp_selection}. 
This implication had also been discussed in \cite{tran2019NCEscore}, in which they theoretically showed that the source task hardness could affect their transferability metric NCE score. However, the extent to which better source models result in better MELEP requires further systematic investigation. In Table \ref{table:full_results}, 
despite underperforming ResNet, Bi-LSTM surprisingly achieved better Pearson correlation coefficients between MELEP and fine-tuning performance. 
This suggests that factors beyond model performance may influence MELEP.
Kornblith et al. \cite{kornblith2019better} might provide valuable insights into this question, as they pointed out that regularization and training settings had an impact on their transferability metric, ImageNet Top-1 Accuracy.
Analyzing the impact of those dependence sources is an interesting topic to explore in the future.

\begin{table}[t]
 \centering
 \def\arraystretch{1.1}%
 \caption{Performance of pre-trained models on the source task (evaluated on the test subset).}
 \label{table:pretrained_performance}
 \begin{tabular}{ccc}
 \hline
 Model & Source Data & Average F1 \\
 \hline
 \multirow{4}{*}{ResNet1d101}
 & PTB-XL & 0.744\\
 & CPSC2018 & 0.740 \\
 & Georgia & 0.593 \\
 & CSN & 0.603 \\
 \hline
 \multirow{4}{*}{Bi-LSTM}
 & PTB-XL & 0.541 \\
 & CPSC2018 & 0.412 \\
 & Georgia & 0.167 \\
 & CSN & 0.277 \\
 \hline

 \end{tabular}
\end{table}

\textbf{Data dependence}: in addition, MELEP is also dependent on the source dataset. 
Equations \ref{core_melep} and \ref{eq_melep} show that the cardinality of the source label set contributes to the MELEP score. 
While MELEP can technically be applied even when the source and target datasets have different label sets, it's reasonable to expect that substantial overlap between the two sets would facilitate transferability. 
Conversely, in cases of minimal overlap, the source model would exhibit greater uncertainty regarding inputs from the target dataset, resulting in a more balanced dummy probability distribution (like random guesses) across non-overlapping diagnostic categories.
For example, suppose that AF (Atrial Fibrillation) is a new label in the target dataset and does not appear in the source one, the uncertainty about AF would result in a dummy probability of positive AF being reduced close to 0.5. Therefore, any negative log-likelihood components related to AF will be larger, leading to a larger MELEP, indicating harder transferability. This effect amplifies with an increased number of non-overlapping categories. Note that here we assume that the source model is good enough, otherwise, even with overlapping labels, the dummy probability distributions may be worse than a random guess, leading to poor transferability predictions.

\textbf{Considerable extensions}: as mentioned in \ref{eq_melep}, we do not consider source label weights in the MELEP formula.
This exclusion is based on the assumption that we lack prior knowledge of the source label distribution used in pre-training.
However, in situations where this information is known, it is more sensible to take the source weights into account.
Additionally, there is another variant that deserves consideration for its practical versatility. Instead of aggregating the weighted average of $\phi (\theta, \mathcal{D}, y, z)$ into a single value as in \ref{eq_melep}, we can output a vector of size $\mathcal{Y}$, indicating the transferability measures for each target label. 
Such an approach is well-suited in scenarios where the performances on certain labels hold more significance than others.

\textbf{Potential applications}: apart from the source checkpoint selection use case demonstrated in Section \ref{subsec:ckp_selection}, MELEP can be useful for continual learning algorithms that based on neural architectures changes or selection of data points in replay buffers \cite{ecg_continual1,ecg_continual2}, facilitating the decision-making process. 
Additionally, in federated learning, where data is often allocated across multiple sources \cite{resnetfed,baumgartner2023introduction} can utilize MELEP to facilitate local model selection and fine-tuning.
Furthermore, multi-task learning \cite{ecg_multitask1,ecg_multitask2}, which often depends on the selection of deep parameter-sharing networks and a combination of task labels, can also benefit from MELEP.
Finally, MELEP holds the potential to assist in the selection of hyperparameters for Bayesian optimization \cite{ecg_bayesian}.
We leave these directions for future work.

\backmatter

\section*{Declarations}

\subsection*{Funding Declaration}
This work was supported by VinUni Seed Grant 2020.

\subsection*{Ethical Approval}
Not applicable.

\subsection*{Competing Interests}
The authors declare no competing interests.


\bibliography{refs}

\end{document}